\documentclass{jkas}

\def\year{2014} 
\def\month{March} 
\def\volume{00} 
\def\issue{0} 
\def\beginpage{1} 
\def\endpage{99} 
\def\received{---} 
\def\accepted{---} 

\date{Received \received ; accepted \accepted}

\title{
{\it WFIRST} Ultra-Precise Astrometry II:  Asteroseismology
}

\author[1]{A.~Gould}
\author[2,3,4]{D.~Huber}
\author[1,5]{M.~Penny}
\author[4,6]{D.~Stello}

\affil[1]{Department of Astronomy Ohio State University,
140 W.\ 18th Ave., Columbus, OH 43210, USA 
\email{gould@astronomy.ohio-state.edu }}
\affil[2]{NASA Ames Research Center, Moffett Field, CA 94035, USA}
\affil[3]{SETI Institute, 189 Bernardo Avenue, Mountain View, CA 94043, USA}
\affil[4]{Sydney Institute for Astronomy (SIfA), School of Physics, 
University of Sydney, NSW 2006, Australia; \email{huber,stello@physics.usyd.edu.au}}
\affil[5]{Sagan Fellow}
\affil[6]{Stellar Astrophysics Centre, Department of Physics and
Astronomy, Aarhus University, Ny Munkegade 120, DK-8000 Aarhus C, Denmark}

\def\corrauthor{
Andrew Gould
}

\def\runningauthor{
Andrew Gould, Daniel Huber, Matthew Penny, Dennis Stello
}

\def\runningtitle{
WFIRST Asteroseismology
}

\def\keywords{
astrometry -- gravitational microlensing  -- stars: oscillations
}

\def\abstracttext{
{\it WFIRST} microlensing observations will return high-precision parallaxes,
$\sigma(\pi)\la 0.3\,\muas$, for the roughly 1 million stars with $H<14$ 
in its $2.8\,{\rm deg}^2$ field toward the Galactic bulge.  Combined with
its 40,000 epochs of high precision photometry ($\sim 0.7$ mmag at 
$H_\vega=14$ and
$\sim 0.1$ mmag at $H=8$), this will yield a wealth of asteroseismic data
of giant stars, primarily in the Galactic bulge but including a substantial
fraction of disk stars at all Galactocentric radii interior to the Sun.
For brighter stars, the astrometric data will yield an external check
on the radii derived from the two asteroseismic parameters, 
$\langle\Delta\nu_{nl}\rangle$
and $\nu_\max$, while for the fainter
ones, it will enable a mass measurement from the single measurable
asteroseismic parameter $\nu_\max$. Simulations based on {\it Kepler} data
indicate that {\it WFIRST} will be capable of detecting oscillations in
stars from slightly less luminous than the red clump to the tip of the red
giant branch, yielding roughly 1 million detections.
}

\def\vega{{\rm vega}}

\def\masyr{{\rm mas}\,{\rm yr}^{-1}}
\def\muas{{\mu\rm as}}

\def\max{{\rm max}}

\def\eff{{\rm eff}}

\def\la{{<\atop \sim}}
\def\ga{{>\atop \sim}}
\def\apj{{ApJ}}
\def\aj{{AJ}}

\def\apjs{{ApJS}}
\def\aap{{A\&A}}
\def\pasp{{PASP}}
\def\mnras{{MNRAS}}
\def\araa{{ARA\&A}}
\begin{document}
\jkashead 


\section{{Introduction}
\label{sec:intro}}

{\it Kepler}'s wide-field, high-precision photometry has revolutionized
asteroseismology. Its $\mu$mag precision at 30 minute cadence
for brighter targets has enabled asteroseismic measurements for
more than 15,000 giant stars, while its special high (1 minute) cadence
feature has even permitted measurements for more than 500
dwarfs.  {\it Kepler} has built on the earlier successes of {\it CoRoT},
which observed a smaller overall number of stars but had the added
advantage of probing several different lines of sight.  See \citet{chaplin13}
for a review.

As a major part of its overall mission, the {\it WFIRST} satellite will
conduct a photometric survey of $2.8\,{\rm deg}^2$ field toward the
Galactic bulge, with roughly 40,000 52s exposures in a broad $H$ band
during six 72-day campaigns at 15 minutes cadence using its 2.4m telescope.
While the main purpose of these campaigns is a search for extrasolar
planets using the gravitational microlensing technique, the survey
will produce a wealth of astrometric and photometric data that can
be applied to a wide range of other astronomical questions.  In 
\citet{gould14} (hereafter Paper I), 
we showed how these astrometric data set could be used to
obtain precise orbits of several thousand Kuiper Belt Objects (KBOs)
down to $H_\vega\sim 28.2$ (corresponding to $R\sim 29.6$)
and how the photometric data could measure thousands of KBO occultations.
These measurements will be concentrated at one extreme of {\it WFIRST}
sensitivity, well below the effective ``sky'' (actually a combination
of true sky, read noise, and dark current) of 341 photons per pixel
per exposure.

Here we consider an application from the opposite extreme,
astrometry and photometry of ``saturated'' stars, i.e., stars
that fill up the full well of the central pixel even in the first
2.6s ``read'' of the 52s exposure.  Because a diffraction-limited
point spread function (PSF) for a circular aperture falls off roughly
as $r^{-3}$,  these ``saturated'' stars still produce excellent
astrometry and photometry.  Indeed, the mission-length parallax precision
$\sigma(\pi)\la 0.3\,\muas$ for about 1 million stars, is likely to
vastly surpass the performance of any other instrument.

By contrast, the photometric precision, roughly 
$\sigma(H)\sim 10^{(2/15)(H-15)}\,$mmag for stars of $8\la H\la 15$
is substantially worse than {\it Kepler}'s, and this for primarily
the same reason that its astrometry is better: whereas {\it Kepler}
deliberately degraded its PSF to spread bright-star photons over
many pixels (Table 1 of \citealt{kepler10}),
{\it WFIRST} will operate near the 2.4m diffraction
limit, thus ``squandering'' most of the photons from the brightest stars.
In addition, {\it WFIRST} will spend only 10\% of its time observing each 
of 10 subfields, but this is basically compensated by its larger mirror
relative to {\it Kepler}.

{\it WFIRST} suffers a second asteroseismic disadvantage 
relative to {\it Kepler}, in addition to worse photometric precision:
it operates in a broad $H$ band rather than {\it Kepler}'s broad visible
band.  Because the amplitudes of stellar oscillations are only about 45\%
as big at $H$ band, this is equivalent to a factor $\sim 2$ further
degradation in precision.

However, {\it WFIRST} partially compensates for these disadvantages
with its high-precision astrometry.  To first order, asteroseismic
measurements of cool stars yield two parameters, the large-frequency separation
$\langle\Delta\nu_{nl}\rangle$ 
and the frequency of maximum oscillation power $\nu_\max$.
The first is a measure of mean density $\rho$ \citep{ulrich86,KB95},
\begin{equation}
{\rho\over\rho_\odot}\simeq 
\biggl({\langle\Delta\nu_{nl}\rangle\over\langle\Delta\nu_{nl}\rangle_\odot}
\biggr)^2,
\label{eqn:rho}
\end{equation}
while the second is a measure of the surface gravity $g$ \citep{brown91,KB95}
\begin{equation}
{g\over g_\odot}\simeq {\nu_\max\over\nu_{\max,\odot}}
\biggl({T_\eff\over T_{\eff,\odot}}\biggr)^{1/2},
\label{eqn:g}
\end{equation}
where $T_\eff$ is the effective temperature.

Obviously, these can be combined to yield the star's radius and mass
\citep{kallinger10}
\begin{equation}
{R\over R_\odot}\simeq 
{\nu_\max\over\nu_{\max,\odot}}
\biggl({\langle\Delta\nu_{nl}\rangle\over\langle\Delta\nu_{nl}\rangle_\odot}
\biggr)^{-2}
\biggl({T_\eff\over T_{\eff,\odot}}\biggr)^{1/2},
\label{eqn:radius}
\end{equation}
and
\begin{equation}
{M\over M_\odot}\simeq 
\biggl({\nu_\max\over\nu_{\max,\odot}}\biggr)^3
\biggl({\langle\Delta\nu_{nl}\rangle\over\langle\Delta\nu_{nl}\rangle_\odot}
\biggr)^{-4}
\biggl({T_\eff\over T_{\eff,\odot}}\biggr)^{3/2},
\label{eqn:mass}
\end{equation}

These ``equations'' (really scaling relations) point to four
related problems in the interpretation of asteroseismic measurements.
First, the relations are approximate and depend as well on evolutionary
stage and chemistry \citep{stello09,white11,miglio13}.
The only way to verify that real masses are
being extracted from Equation~(\ref{eqn:mass}) is to apply it
to rare test cases of stars with masses that are measured from binary-star
orbits\footnote{See \citet{epstein14} for a less stringent but still
interesting such test.} \citep{sandquist13,frandsen13},
or to apply the companion Equation~(\ref{eqn:radius}) for
stars with known radii \citep{huber12,white13}.  
In the Gaia era, there will be a large number
of giant stars in the solar neighborhood (within 1 kpc) 
with accurate ($\la 1\%$) parallaxes, and so accurate radii based 
on the well measured infrared color/surface-brightness relations.
This includes several hundred giants with {\it Kepler} asteroseismology,
which will therefore provide crucial checks on these scaling relations,
as well as calibrations of the deviations from them based on various other
observables.  However, it will not be easy to extend these
calibrations based on ``garden variety'' solar-neighborhood stars to the
much more extreme populations found in the Galactic bulge.

Second, errors in the observed quantities $\nu_\max$ and 
$\langle\Delta\nu_{nl}\rangle$
are amplified by factors of three and four respectively before they
enter the mass.  Thus even if the deviations from the scaling relations
are understood perfectly, it can be difficult to extract masses
(hence ages) of individual stars.

Third, the assumption underlying the $\nu_\max$ scaling relation proposed
by \citet{brown91} (a linear relation of $\nu_\max$ with the acoustic
cut-off frequency) is theoretically less understood than the scaling
of $\langle\Delta \nu_{nl}\rangle$, 
potentially introducing unknown systematic errors in mass estimates
based on Equation (4). While theoretical work explaining the $\nu_\max$
relations has yielded some progress (e.g.,\citealt{belkacem11}), mass
measurements based on $\langle\Delta \nu_{nl}\rangle$
should therefore yield results that we can more comfortably interpret.

Fourth, if the data are sufficiently noisy, it will be possible to
measure only $\nu_\max$ (and not $\langle\Delta\nu_{nl}\rangle$), 
which would yield
only a surface gravity but not a mass.

All four of these problems can be solved or mitigated by {\it WFIRST}
parallaxes, which yield stellar radii (assuming the surface brightness
can be properly estimated).  First, for brighter stars, which have
both better photometric precision and larger amplitudes, it will
be possible to directly check the radius derived by
applying Equation~(\ref{eqn:radius}) to asteroseismic measurements.
Second, if the radius $R$ is known independently, then the mass can be written
\begin{equation}
{M\over M_\odot}\simeq \biggl({\langle\nu_{nl}\rangle\over
\langle{\nu_{nl}\rangle_\odot}}\biggr)^2
\biggl({R\over R_\odot}\biggr)^{3}
\label{eqn:mass2}
\end{equation}
\begin{equation}
{M\over M_\odot}\simeq {\nu_\max\over\nu_{\max,\odot}}
\biggl({T_\eff\over T_{\eff,\odot}}\biggr)^{1/2}
\biggl({R\over R_\odot}\biggr)^{2}.
\label{eqn:mass3}
\end{equation}
For higher S/N detections, Equation~(\ref{eqn:mass2}) removes the
dependency of the mass estimate on the less well-tested $\nu_\max$ scaling
relation (and so permits precise testing of this relation), while
for low S/N detections, Equation~(\ref{eqn:mass3}) still permits
a mass measurement (whose accuracy will be determined by the tests
just mentioned).

Hence, {\it WFIRST} astrometry can transform relatively crude {\it WFIRST}
asteroseismic measurements into precision mass and 
(when metallicities are available) age measurements.

\begin{figure}
\centering
\includegraphics[width=90mm]{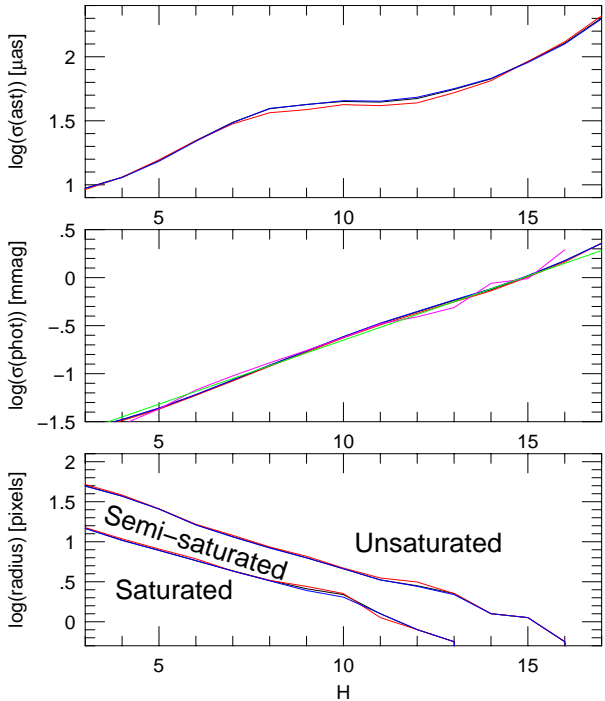}
\caption{Astrometric (top) and photometric (middle) precision from
a single-epoch of simulated {\it WFIRST} observations as a function
of $H$-band apparent magnitude.  Red, black and blue are for
thermal spectral energy distributions at 2500, 4500, and 6500 K.
In the middle panel, the magenta curve shows the analytic calculation
(Eq.~(\ref{eqn:sntotphot3})) using $r_{\rm unsat}$, calculated
numerically from $\pi r_{\rm unsat}^2= N_{\rm sat} + N_{\rm semi-sat}$
(total number of saturated and semi-saturated pixels) as displayed
in the bottom panel (upper curves).  The agreement is excellent.
Also shown in green is a pure power law 
$\log\sigma({\rm phot}) = (2/15)H +const.$  The approximately
flat astrometric errors $\sigma({\rm ast})\sim 60\,\muas$ (top panel)
are predicted for an Airy disk when $r_{\rm sat}>0$ (bottom panel).
However, this relation holds only for $H\ga 8$, brighter than which
the diffraction spikes start to contribute to the precision.
}
\label{fig:prec}
\end{figure}

\section{{Astrometric and Photometric Precision Saturated Stars 
in the Photon-Noise Limit}
\label{sec:saturated}}

The first of many steps toward estimating the precision of parallaxes
and asteroseismic parameters is to establish the precision of individual
measurements in the photon-noise limit.  Of course, since the
parallax precisions of $0.3\,\muas$ claimed above are $10^{5.6}$
smaller than the {\it WFIRST} pixel size, it is clear at the outset that
even very subtle systematics may be the ultimate limit of these measurements.
Nevertheless, a clear understanding of the statistical errors is an essential
prerequisite for a discussion of systematics.

Figure~\ref{fig:prec} shows the results of a Fischer matrix calculation of the
astrometric and photometric errors.  We begin with a $2048\times 2048$
representation of the PSF for monochromatic $1.0\,\mu$m light, with
11 mas pixels, i.e., 10 times smaller than the {\it WFIRST} pixels.
We then convolve this with several different spectral energy distributions,
with the PSF scaled by wavelength.  We find numerically
that the results do not vary significantly
for radically different spectral energy distributions.  We assume
(as in Paper I) 52s exposures, a photometric zero point of $H_\vega=26.1$,
and 341 photo-electrons per pixel in sky, read noise, and dark current.
We assume that the full well has a depth of 100,000 photo-electrons,
and that it is read in 20 equal 2.6s ``reads''.  Hence, for example,
if 52s of flux would generate 2,000,001 photo-electrons, the pixel
would be saturated; for 1,999,980 electrons a single read would record
99,999, so the fractional photon-noise error would be 0.32\%; for
1,000,020 electrons, a single read would record 50,001, so the fractional
error would be 0.45\%, etc.  In spite of this somewhat irregular pattern,
the basic effect of this read structure is that the fractional errors are
near 0.32\% over the entire range $1\times 10^5<N<2\times 10^6$.  At higher
fluxes there is no information while at lower fluxes the errors increase
as $1/\sqrt{N}$ in the usual way.

We explicitly assume two things that are known not to be true.
First, we assume that the sensitivity of the pixel is spatially uniform.
Second, we assume that the detector response is linear below saturation.
These assumptions have no practical impact on the
signal-to-noise calculations, although they will have to be carefully
taken into account in the actual measurements.  

The astrometric covariance matrix is then given by
\begin{equation}
c = b^{-1}, \qquad b_{i,j} = (\delta\theta)^{-2}
\sum_k {(\delta \ln F_{k,i})(\delta \ln F_{k,j})\over [\sigma(\ln F_k)]^2}
\label{eqn:astmat}
\end{equation}
where $\delta \ln F_{k,i}$ is the fractional flux change when pixel $k$
is displaced in the $i$ (either $x$ or $y$) direction by an angle
$\delta\theta$ (in practice 0.1 pixels, i.e., 11 mas), and
$\sigma(\ln F_k)$ is the fractional photon error, discussed immediately above.
The fractional photometric error is simply $[\sum_k \sigma(\ln F_k)^{-2}]^{-1/2}$.

When discussing systematics, it will be important to understand the
origin of the principal features of Figure~\ref{fig:prec}.  The photometric
error is an essentially featureless power law, $\sigma(\ln F)\propto F^{-1/3}$,
which appears as $\log\sigma(\ln F)=(0.4/3)(H-15)\sim 0.13(H-15)$ on the plot.
By contrast, the astrometric error is essentially flat over $8<H<14$, which
is the principal range of interest for the current work.  At both fainter
and brighter magnitudes, it approaches slightly different power-law
slopes of $-1/2$ and $-1/3$, respectively.

The {\it WFIRST} point spread function (PSF) is basically comprised
of two components: an Airy disk and 12 diffraction spikes.
The main features of Figure~\ref{fig:prec} can be understood analytically from
these two components.

\subsection{{Astrometric Precision of Saturated Broad Band Airy Disk}
\label{sec:airy}}

The diffraction pattern due to a circular mirror of diameter $D$ for 
monochromatic light of wavelength $\lambda$
is an ``Airy disk'',
\begin{equation}
I = 4 I_0\biggl({J_1(x)\over x}\biggr)^2;
\qquad x = \pi{D\over \lambda}\sin\theta;
\qquad I_0 = P_0\biggl({\pi D\over 2\lambda}\biggr)^2
\label{eqn:airy}
\end{equation}
where $I$ and $I_0$ are intensity in power per steradian, $J_1$ is
a Bessel function, and $P$ the total power incident on the mirror.
The asymptotic behavior of the Bessel function is 
$J_1(x)\rightarrow (\pi x/2)^{-1/2}\cos(x - 3\pi/4)$.   Hence,
the Airy disk scales $I\sim \cos^2(x-3\pi/4)/x^3$.

We now focus on applications to the very broad {\it WFIRST} wide
$H$-band filter, which covers a factor $\sim 2$ in wavelength,
The fourth zero $(x=13.324=4.21\pi$) of the central
wavelength ($\lambda_0=1.5\mu$m) at $\theta=542\,$mas, corresponds
a range of $6.36>x/\pi>3.18$ for wavelengths 
$(2/3)\lambda_0,\lambda<(4/3)\lambda_0$. Hence, even at a few
Airy lengths, the combined flux per pixel from all wavelengths is well
approximated by
\begin{equation}
F(r) = {k\over r^\gamma};\qquad \gamma=3
\label{eqn:fpix}
\end{equation}
where $r$ indicates the distance of the pixel from the center
in pixels and $k$ is a constant.  Of course, there will be
some residual wavy structure superposed on this power law, and
this will have to be taken into account in the actual measurement.
However, from the standpoint of determining the astrometric precision
this is unimportant.  Moreover, to the extent that this structure
enters, it {\it adds} information, and therefore ignoring it is
conservative.  More generally, we can consider an arbitrary power law
$r^{-\gamma}$.

Let us now consider that the well has a capacity for $n_\max=10^5$ 
photo-electrons and that there are $N_{\rm read}=20$ reads during the
exposure, so that pixels with more than $n_\max N_{\rm read}=2\times 10^6$
photo-electrons are fully saturated (no information).  
We define $r_1$ as the radius
at which $n_\max$ photo-electrons are captured in a single read time,
and $r_m=m^{1/\gamma}r_1$ for $m\leq N_{\rm read}$.  We call the region
$r_1<r<r_{N_{\rm read}}=N_{\rm read}^{1/\gamma}r_1$ ``semi-saturated'', since
it is unsaturated for some $m<N_{\rm read}$ 
reads but saturated in $N_{\rm read}$ reads.
Then for pixels at 
$r_m<r<r_{m+1}$, a total of $(r/r_m)^{-\gamma}n_\max$ photo-electrons will
be read during $m$ reads, which lies in the range $(m/(m+1),1)n_{\rm max}$.

If the source is displaced along the $x$-axis by $\epsilon$ pixels, then
this leads to a change in flux $(d\ln F/d\ln r)\epsilon$ compared
to a fractional error $[n(r)]^{-1/2}$, and hence a signal-to-noise ratio
for this one pixel of
\begin{equation}
\biggl({\rm S\over N}\biggr)_i^2 = \biggl({\gamma\over r}\epsilon\biggr)^2
\biggr({r\over r_m}\biggr)^{-\gamma}n_\max
\label{eqn:sn1}
\end{equation}
Summing over all pixels in the semi-saturated range, taking account
of the fact that pixels lying off the $x$-axis
contribute to $\rm (S/N)^2$ as $\cos^2\phi$,
and turning the sum into an integral, we obtain,
\begin{eqnarray}
\biggl({\rm S\over N}\biggr)_{\rm semi-sat}^2 &=& 
\sum_i \biggl({\rm S\over N}\biggr)_i^2 
\nonumber \\ &\rightarrow&
\pi\gamma^2\epsilon^2 n_\max \sum_{m=1}^{N_{\rm read}-1}
\int_{r_m}^{r_{m+1}} {dr\over r}
\biggr({r\over r_m}\biggr)^{-\gamma}
\nonumber \\
&=&\pi \gamma \epsilon^2 n_\max \sum_{m=2}^{N_{\rm read}} {1\over m}
\label{eqn:sntot}
\end{eqnarray}
A similar calculation for the region just beyond the saturation zone
gives 
\begin{equation}
\biggl({\rm S\over N}\biggr)^2_{\rm unsat} = \pi\gamma\epsilon^2 n_\max.
\label{eqn:snunsat}
\end{equation}
Hence the error in
estimating this offset (e.g. \citealt{gould97}) is
\begin{eqnarray}
\sigma(\theta) &=& {p\over 
\sqrt{\pi\gamma n_\max\ln (1.78 N_{\rm read} + 0.9)}}
\nonumber \\ &=& 60\muas 
\biggl({p\over 110\,\rm mas}\biggr)
\biggl({n_\max\over 10^5}\biggr)^{-1/2}
\nonumber \\ &\times&\biggl({\ln (1.78 N_{\rm read}+0.9)\over 3.6}\biggr)^{-1/2}
\biggl({\gamma\over 3}\biggr)^{-1/2}
\label{eqn:airysat}
\end{eqnarray}
where $p$ is the pixel size\footnote{
We use $\sum_{m=1}^N m^{-1}\simeq \ln(1.78N + 0.9)$ rather than the limiting
Euler formula because it is accurate for low $N$, down to $N=1$.}.

Note that this expression is completely independent of the source
brightness: it only requires that the source is saturated.  
This explains the flat behavior of the astrometric precision in the
range $8<H<14$ in Figure~\ref{fig:prec}.  In the range  $13<H<16$, the central
pixel is unsaturated in a single 2.6s read and the semi-saturated region
gradually disappears, yielding a 
gradual transition to standard ``root-N'' behavior.  On the other hand,
at brighter magnitudes $H<8$, the diffraction spikes become important 
and add a new source of ``signal''.   In contrast to the Airy disk,
the effective width of the diffraction spikes does not increase
with source brightness, so their information content grows directly
with source brightness, leading to improved astrometric precision.
At the brightest magnitudes $H\la 4$ (which are not of practical interest),
the finite $(2048\times 2048)$ size of the simulated PSF cuts off
important regions of astrometric information, which induces a spurious
flattening.

\subsection{{Photometric Precision of Saturated Broad Band Airy Disk}
\label{sec:airy_phot}}

A similar calculation for photometric precision yields
\begin{eqnarray}
\biggl({\rm S\over N}\biggr)_{\rm semi-sat,phot}^2 
&\rightarrow& 2\pi n_\max \sum_{m=1}^{N_{\rm read}-1}
\int_{r_m}^{r_{m+1}} dr\,r
\biggr({r\over r_m}\biggr)^{-\gamma}
\nonumber \\  = {2\pi n_\max\over \gamma -2} &r_1^2&
\biggl(-N_{\rm read}^{2/\gamma}+\sum_{m=1}^{N_{\rm read}} m^{2/\gamma -1}
\biggr),
\label{eqn:sntotphot}
\end{eqnarray}
which can be expressed
$$
\biggl({\rm S\over N}\biggr)_{\rm semi-sat,phot}^2 
= \pi n_\max r_{\rm unsat}^2(1 - Q); \nonumber
$$
\begin{equation}
Q\equiv
{N_{\rm read}^{-1} +\gamma (2N_{\rm read})^{-2/\gamma}
\over \gamma -2}
\label{eqn:sntotphot2}
\end{equation}
where $r_{\rm unsat} = N_{\rm read}^{1/\gamma}r_1$ is the smallest radius
at which the full readout is unsaturated. A trivial calculation
yields ${\rm (S/N)^2_{\rm unsat, phot}} = 2\pi n_\max(\gamma-2)^{-1} r_{\rm unsat}^2$.
Hence the fractional photometry error is
\begin{eqnarray}
\sigma (\ln F) &=& \biggl[\pi n_\max 
\biggl({\gamma\over\gamma-2} - Q\biggr)\biggr]^{1/2}r_{\rm unsat}^{-1}
\nonumber \\
&=& 2.1\times 10^{-4}
\biggl({n_\max\over 10^5}\biggr)^{-1/2}
\nonumber \\ &\times& 
\biggl({\gamma/(\gamma-2)-Q\over 2.7}\biggr)^{-1/2}
\biggl({r_{\rm unsat}\over 5}\biggr)^{-1}
\label{eqn:sntotphot3}
\end{eqnarray}
Note that since $r_{\rm unsat}\propto F^{1/\gamma}$ the fractional
photometry precision scales the same way, i.e., as $F^{1/3}$ for
a broad-band Airy disk.  The magenta curve in the middle panel
of Figure~\ref{fig:prec} is a numerical evaluation of the rhs
of Equation~(\ref{eqn:sntotphot3}), with $\gamma=3$ and
$r_{\rm unsat}$ evaluated numerically in the top panel
from $\pi r_{\rm unsat}^2 = (N_{\rm sat} + N_{\rm semi-sat})$, i.e., the
total number of saturated and semi-saturated pixels.  The agreement
is essentially perfect.

Finally, we note that semi-saturated annulus dominates the
astrometric information while the unsaturated outer regions
dominate the photometric information.  However, in neither
case is this dominance overwhelming.  For the astrometric precision,
the ratio (semi-saturated/unsaturated) is 2.6:1, whereas for the
photometric precision it is 1:2.9.  This will have important
implications for controlling systematics.

\subsection{{Utility of Analytic Formulae}
\label{sec:util}}

The fact that the curves in Figure~\ref{fig:prec} can be understood
qualitatively and quantitatively from analytic formulae allows one
to quickly estimate the impact of a wide range of changes in system
parameters.  For example, at this point it is undecided whether
{\it WFIRST} will have 2.6s reads or 5.2s reads.  If the latter,
then $N_{\rm read}\rightarrow 10$ (rather than 20).  
Equation~(\ref{eqn:airysat}) then immediately implies a
degradation of astrometric precision of $(\ln 36.5/\ln 18.7)^{1/2}=1.11$.
The estimate of the change in photometric precision requires one 
more step.  Adopting $\gamma=3$, we find that $Q=0.51$ (rather than 0.31).
Hence, the photometric precision is degraded by 
$((3-0.31)/(3-0.51))^{1/2} = 1.04$.

\section{{From Astrometric Measurements to Parallaxes}
\label{sec:ast_par}}

Even assuming that there were no errors in single-epoch astrometry
other than those due to photon noise, there are still several steps
required to go from the ensemble of such measurements to precision
parallaxes.  We outline these steps as a preliminary to discussing
systematic errors.

First, each astrometric measurement is made relative to the pixel
grid of the detector, but what is of practical interest is the
position of the star relative to a frame of reference set by the
sky.  Because the observations will be dithered, perhaps over a
total range of $20^{\prime\prime}$, the transformation from pixel-grid
frame to sky frame must be made for each image.  The only sky ``object''
whose position can be known as well as the very bright stars that
are the subject of this paper, is some ensemble of fainter stars
that are near-enough (in angle) that their measured pixel separations
can be securely translated into angular separations.  The choice
of this ``ensemble'' is crucial because in the next step, the
absolute parallax of this ``object'' will have to be measured
extremely precisely in order to translate the ``relative parallax''
described just below (between the ``bright star'' and the 
``ensemble object'') into an absolute parallax.

With 400,000 (40,000 epochs for each of 10 fields)
dithers over $\sim 20^{\prime\prime}\times 20^{\prime\prime}$, the
{\it relative} offset between pixels (on some definite, but so
far unknown scale) will be determined extremely precisely over these
separations by the fact that the great majority of these stars are moving
relative to each other only according to their relative parallax,
proper motion, and, possibly, modeled binary-motion orbits.
That is, there will be of order one star per square arcsec, with 
individual-epoch astrometric precision of $<1\,\muas$.  Hence, from
these $(4\times 10^7)\times(4\times 10^5)\sim 10^{13}$ it will be
quite straightforward to constrain the relative offset of pixels,
including terms for time evolution and for color and well-filling.
This will not, by itself, establish an absolute angular scale,
a topic to which we return below.  However, the fractional error
in the derived parallax due to an inaccurate angular scale is simply
equal to the fractional error in the scale, which for bulge stars
is orders of magnitude below the parallax statistical errors.

The reference ensemble should then be chosen to have 
similar parallaxes (in $\muas$, not distance modulus), 
but selected primarily on something other than
measured parallax in order to minimize bias.  An excellent choice
would be stars whose color and magnitude are consistent with
membership in the bulge red clump (RC) and whose proper motions are consistent
with being in the bulge (e.g., with proper motions relative to the mean
of bulge stars no greater than $4\,\masyr$).  There will be relatively
few foreground contaminants to such a sample, and these can be eliminated
using {\it WFIRST} relative parallaxes.

Unfortunately, the total number of RC stars within, say $20^{\prime\prime}$
is far too small to measure the mean absolute parallax with the
requisite precision.  The density of clump stars in the {\it WFIRST}
microlensing fields is 1--$2\times 10^{5}\,{\rm deg}^{-2}$, meaning that
only 1000--2000 lie within this radius.  This may sound like a lot,
but in these heavily extincted fields, the RC is at Gaia magnitudes
of $18\la G \la 20$ (or fainter in some cases), implying Gaia parallax 
precisions\footnote{http://www.cosmos.esa.int/web/gaia/science-performance}
of $100\,\muas\la\sigma(\pi)\la 300\,\muas$, and so ensemble precisions
of $3\,\muas\la\sigma(\pi)\la 10\,\muas$, which is impressive but still
a factor 15--50 times larger than the {\it WFIRST} individual relative parallax
errors for the brightest stars and 3--10 times larger than the $1\,\muas$
(so $\la 1\%$) parallax errors relevant to the precision radius measurements
needed for asteroseismology.

However, one could consider the ensemble of all RC stars over 
the {\it WFIRST} fields and fit the difference between local {\it WFIRST}
relative parallaxes and absolute Gaia parallaxes to a linear (or quadratic)
function.  In principle, such an approach runs the danger that
discrete (so, non-linear) structures in the bulge would systematically
corrupt the fit, but such structures (if they existed) would easily
show up in the highly precise {\it WFIRST} relative parallaxes alone.

\section{{Systematic Errors}
\label{sec:systematics}}

Systematics can potentially degrade and/or undermine the measurement
process outlined in  Sections~\ref{sec:saturated} and \ref{sec:ast_par} 
in two distinct ways.  First, there can be systematic errors that
corrupt the individual measurements, but in a way that is not
correlated from measurement to measurement.  These effectively
add (in quadrature) to the statistical errors.
Hence, they only need to be controlled at the level of the photon noise
of individual measurements.  For example, for $H=15$ stars (near the RC),
the astrometric errors must be controlled at $\la 100\,\muas$ and the
photometric errors at $\la 1\,$mmag.  On the other hand purely correlated
errors must be controlled at levels that are $\sqrt{N}=200$ times
smaller, where $N\sim 40,000$ is the number of independent measurements.
That is, for $H=15$, at the $0.5\,\muas$ level.  Physically, this
corresponds to about one Bohr radius on the detector.  Even developing
an intuition about what can go wrong at this level is challenging.

\subsection{{Random Systematics}
\label{sec:random}}

We begin with the first issue, uncorrelated systematics.  For example,
the calculations leading to Figure~\ref{fig:prec} were made under the assumption
that the detector response was uniform across the pixel.  This, of
course, will not actually be true.  However, all that is really
required is that the detector response be known as a function of
position, not that it be uniform.  This position-dependent response
will in fact be measured very well.  In any given exposure, roughly 10\% of
pixels will receive flux from stars at a level larger than the ``sky''
background, and the flux from these stars, as well as the position
of the star will be known almost exactly from the ensemble of
40,000 exposures from each field.  Hence, each pixel will be
subjected to $400,000\times 10\% = 40,000$ random flux experiments
with stars with a broad range of colors and brightness.  The
ensemble of these, over $3\times 10^8$ pixels will be used to
measure the mean response, so that the 40,000 flux experiments will
really only be used to find the deviations from the mean and the time
evolution of individual pixels.  These same flux experiments will
measure pixel response as a function of well-depth and mean wavelength
of incident flux.  Moreover, as mentioned at the end of 
Section~\ref{sec:saturated}, there will be roughly equal information 
coming from the semi-saturated and unsaturated pixels, which permits
internal tests on each individual astrometric or photometric measurement,
and which allows to test for extremely subtle deviations from the
ensemble of such differences.  Similarly, one can search for differences
according to angular position of pixels with respect to the center.

Hence, while it is not possible to enumerate in advance all conceivable
sources of uncorrelated systematic errors, it will be straightforward
to quantify the level of such errors (from excess noise they create)
and to construct
many possible tests to locate these errors and calibrate them out.

\subsection{{Correlated Systematics}
\label{sec:correlated}}

{\it WFIRST} microlensing observations will be carried out in six
72-day campaigns, centered on the equinoxes (since the microlensing
fields are near the winter solstice).
Correlated errors in the parallax measurement will occur if the position
of the target star is shifted relative to the ensemble of comparison
stars differently in the spring campaigns than the fall campaigns.
As emphasized above, effects leading to shifts of one Bohr radius,
or even smaller, must be considered.

The two main avenues for such systematics effects are first that the
camera is rolled $180^\circ$ between equinoxes and second that the Sun
is on the opposite side of the field (but the same side of the camera).
One can easily imagine that these factors can lead to systematic
effects of some sort at the Bohr-radius level.  However, it is important
to bear in mind that these effects, whatever they are, must operate
differently on the target star compared to the ensemble of reference stars.
It cannot, for example have anything to do with the particular
pixel that the target star lands on, because it lands on 20,000 different
random pixel positions in each of the spring and fall configurations.
There are only two properties that are systematically different: brightness
and color.

As already mentioned, it is straightforward to test for brightness
dependence by comparing the positions (and so parallaxes) derived
from the semi-saturated versus unsaturated pixels.  Since there
are of order $10^6$ stars on which this test can be performed,
extremely small effects as a function of other control variables
(like proximity to the Sun-side of the detector) can be detected.
Thus, there are good prospects not only for detecting and measuring
such effects, but also determining their origin and thus calibrating
them out.  

One might think of statistically comparing the mean
derived parallax as a function of star brightness, but there is an
obvious source of contamination to such a test: brighter stars
might be systematically nearer.  However, one can compare the
mean parallax difference between {\it WFIRST} relative parallaxes
and Gaia absolute parallaxes as a function of brightness, which
is immune to such bias.  Similar tests could be done as a function
of source color (which is highly correlated with brightness).

Thus while we cannot prove in advance that all such systematics
can be controlled, there are at least good prospects for detecting
them and, most likely, identifying the source and removing them.

\subsection{{Angular Radii}
\label{sec:angular}}

To determine the physical radius $R$ using a trigonometric-parallax-based
distance $d$, one must independently determine the angular radius,
$\theta_*$, i.e., $R=d\theta_*$.  This in turn requires knowledge
of the star's surface brightness and dereddened flux in the same
band.  Depending on the precision required, determining these
parameters can be non-trivial.  The only systematic effort to measure
such angular radii from colors and magnitudes
in significantly reddened fields of which we
are aware is in gravitational microlensing studies, for which the
claimed precision (when data are of good quality) is 7\% 
(e.g., \citealt{gould14b}).  We first review the established procedure
and then outline why one may expect significant improvements
for {\it WFIRST} targets.

As discussed by \citet{ob03262}, the mathematical model of the microlensing
event yields the instrumental magnitudes of the source (free of blending)
in several bands, usually including $V$ and $I$.  One finds the offset
of these values $\Delta ((V-I),I)$
from the RC centroid, using the same instrumental
photometry.  The dereddened color of the clump is known to be 
$(V-I)_{0,\rm cl}=1.06$ from the work of \citet{bensby13}, 
which we briefly recount
below, while the dereddened magnitude $I_{0,\rm cl}$ is known as a function
of field position from the study by \citet{nataf13}.  These yield
$(V-I,I)_0 = (V-I,I)_{0,\rm cl} + \Delta ((V-I),\Delta I)$.  Then this
$V/I$ photometry is converted to $V/K$ using the empirical color-color
relations of \citet{bb88}.  The $(V-K)_0$ color is used to estimate
the $K$-band surface brightness employing the empirical color/surface-brightness
relations of \citet{kervella04}, and finally this is combined with $K_0$
to determine the source angular radius, $\theta_*$.  
(See \citealt{boyajian14}, for an update of these relations.)

There are two
principal sources of uncertainty in this estimate.  First, the dereddened
color is determined only to about $\sigma(V-I)_0\sim 0.05$ mag.  This
uncertainty is known because the color-estimation procedure has been applied by
\citet{bensby13} to a sample of about 50 dwarfs and subgiants with
high-resolution spectra (taken when the source was highly magnified
by microlensing).  Then the $(V-I)_0$ colors were predicted from
models based on spectral classification and compared to those
determined by the microlensing method.  For relatively blue stars
near the turnoff, the scatter is about 0.06 mag, of which some contribution
is due to the uncertainty in the spectroscopic temperature, implying
that the intrinsic scatter in the microlensing method is 0.05 mag
(or possibly less, if there are other unrecognized errors in the
spectroscopic determinations).  Redder microlensed stars show greater
scatter but \citet{bensby13} argue that this is due to uncertainty
in the spectroscopic models of these stars.  Note that \citet{bensby13}
determine the color of the RC by choosing the value that minimizes
this scatter.

Second, there is typically a 0.1 mag uncertainty in estimating the $I$-band
magnitude of the
RC centroid.  These two errors combined yield a 7\% error in $\theta_*$.
No account is usually taken of errors in the overall distance scale
(i.e., $R_0$) derived by \citet{nataf13} nor in the 
color/surface-brightness relations derived by \citet{kervella04}, since
these are deemed small compared to the dominant errors.

The angular radius estimates for giant stars observed by {\it WFIRST}
will follow the broad outline of microlensing estimates but differ
in many particulars.  Most important, the main flux measurement will be
carried out in the infrared, probably $K$ band, to enable direct
comparison with the most stably calibrated color-surface brightness
relation.  This removes the use of the $VIK$ color-color relations
of \citet{bb88}, which are in fact truly valid only for near-solar
metallicity stars.

This will also permit a much longer color baseline for determining
the color of RC stars, either in $V-K$ or $I-K$ (for more heavily
reddened fields).  Note that for these very bright stars, ground-based
surveys can provide good $V$ and/or $I$ band photometry,
given that {\it WFIRST}'s own star catalog will permit excellent
subtraction of fainter blends in the great majority of cases.

Second, the depth of the RC will be almost completely removed
by the relative-parallax measurements coming from {\it WFIRST} itself.
Of course, the intrinsic ``height'' of the RC will
remain, which has been estimated by \citet{nataf13} to be
$\sigma(M_{I,\rm cl})=0.09\,$mag, which is 3--4 times smaller
than the height dispersion in typical current microlensing fields.  
This will not only improve the
precision of the RC centroid in the vertical direction, it will
permit more restrictive identification of RC stars and (by combining
deviations from the centroid in both color and magnitude) more secure
identification of RC stars affected by differential extinction.
Of course, since {\it WFIRST} fields will typically have higher extinction
than current microlensing fields, they will also have higher differential
extinction at fixed angular scale. {\it This is important because 
differential extinction induces a difference between the extinction
estimate from an ensemble of clump stars in the neighborhood of a given
star and the true extinction toward that star\footnote{Fortunately, this
leads to two effects that mostly cancel.  That is, underestimating
extinction causes one to underestimate the true source brightness, and
so underestimate its radius.  However, it also causes one to underestimate
the temperature, and so the surface brightness, and hence overestimate
the radius.}.}
However, the density of RC stars
will also be higher, and since the RC (corrected for distance) is tighter,
much smaller angular scales need to be probed to establish the RC centroid.

Thus, if the color/surface-brightness relations can be well-understood
(see discussion below and \citealt{gould14b}),
there are good prospects for measuring angular radii to several times
better than the $\sim 7\%$ achieved 
in current microlensing experiments, i.e., $\sim 2\%$.

We now turn to the two sources of uncertainty that are presently
ignored because they are too small relative to the color and magnitude
errors: uncertainty in $R_0$ and uncertainty in the color/surface-brightness
relations.  The first will be all but eliminated by the combination
of Gaia and {\it WFIRST}.  Gaia by itself will determine the mean
distance to bulge RC stars with great precision and {\it WFIRST} will
tie individual RC stars to this system at extremely high precision.

The second problem is much more severe.  The \citet{kervella04}
color/surface-brightness relations are very tight but the
calibration is based on interferometry of stars that are (necessarily)
in the solar neighborhood and hence primarily of near-solar metallicity.
However, since $V$-band is also sensitive to metallicity through
line blanketing, it is far from clear that the empirical relations
will apply to some of the extreme stars in the Galactic bulge.
To generate a 1\% error in $\theta_*$, the $V$ magnitude must be ``wrong''
(relative to a solar-metallicity star of the same $K$-band
surface brightness) by only 0.065 mag.

This problem is therefore actually two-fold.  First, to be applied
to the extremely metal-rich sub-populations that inhabit the
Galactic bulge, the relations must be calibrated as a function of
metallicity.  Second, if these relations are found to depend on
metallicity, then the {\it WFIRST} stars to which it is applied must have
metallicity measurements (or they will have increased uncertainty due to lack
of knowledge of the metallicity).

The problem of calibration is more fundamental.  As discussed in 
detail by \citet{gouldcal14}, there are extremely few local calibrators
that would be accessible to the workhorse techniques of lunar occultations
and long baseline interferometry.  He therefore develops a new
microlensing-based technique for calibrating the color/surface-brightness
relation for metal-rich stars.  In essence, for a subset of microlensing
events it is possible to measure the source radius crossing time,
i.e., $t_*\equiv \theta_*/\mu$, where $\theta_*$ is the source radius
and $\mu$ is the lens-source relative proper motion.  One then waits
a sufficient time $\Delta t$ for the source and lens to have a measurable
separation $\Delta\theta$, and then derives $\mu=\Delta\theta/\Delta t$,
and so $\theta_* = \mu t_*$.  There are a number of technical challenges
to making these measurements, but they do appear to be feasible.
We refer the reader to \citet{gouldcal14} for details.

There are approximately 1 million stars with $H<14$ in the {\it WFIRST}
fields.  We show below that this is a conservative boundary for
stars with useful astroseismic data.  Even with the already existing
APOGEE infrared spectrograph, which has 300 fibers, mounted on, e.g.,
the 2.5m DuPont telescope, it would be possible to obtain about
600 (S/N=50) spectra per night, which is sufficient to determine
bulk metallicities.  Hence, of order 5\% of the full sample
could be measured in a single season, 
including the great majority of the brighter targets, which
can be observed with much shorter exposures.
This would be sufficient to determine what
additional level of investment was warranted by the asteroseismic
data.  By the time that {\it WFIRST} is launched, more ambitious
infrared spectrographs may be in operation.  For example, the
MOONS\footnote{http://www.roe.ac.uk/~ciras/MOONS/Overview.html} 
spectrograph \citep{moons}
proposed for the 8.2m VLT would have 1000 fibers
over 500 arcmin$^2$.  It could obtain S/N=50 spectra for 1 million
targets $H<14$ in a dedicated 40 day campaign.

\section{{Simulated {\it WFIRST} Asteroseismology}
\label{sec:simul}}

Our basic approach to simulating {\it WFIRST} asteroseismology
is to take real {\it Kepler} lightcurves and add noise.
This implicitly treats {\it Kepler} lightcurves as noise-free,
which of course is not strictly true, but is appropriate because
{\it WFIRST} noise is much larger than {\it Kepler} noise.  Hence,
the difference between this and a ``correct'' approach is much
smaller than the uncertainty in estimating the {\it WFIRST} noise.

\begin{figure}[ht!]
\begin{center}
\resizebox{85mm}{!}{\includegraphics{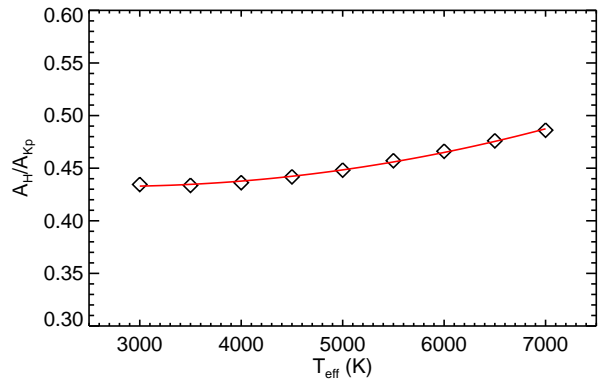}}
\caption{Ratio of $H$-band to $Kp$-band amplitude as a function of 
$T_\eff$ assuming  blackbody curves and 
that the luminosity variations are entirely due to changes in temperature. 
The red solid line shows a second order polynomial fit to the data
(Equation~\ref{eqn:parab}).}
\label{fig:ampratio}
\end{center}
\end{figure}

There are two additional adjustments that we must make to the
{\it Kepler} data beyond adding {\it WFIRST} noise.  First,
we must change the amplitude of the stellar oscillations due to
the fact that these are less pronounced in the {\it WFIRST} $H$-band
than in the {\it Kepler} band.  Second, we must 
account the seasonal gaps in the {\it WFIRST} data that do not occur in
{\it Kepler} data, since these can introduce aliasing.  
In addition, we must account for the fact that
the {\it WFIRST} cadence is twice as fast as the {\it Kepler} cadence
(15 vs.\ 30 min), but this we do simply by decreasing {\it WFIRST}
errors by $\sqrt{2}$.

{\subsection{Amplitude Adjustment}
\label{sec:ampl}}

We begin by assuming that brightness oscillations are purely due
to temperature changes.  In principle, radius variation also
play a role but these are very small by comparison
\citep{KB95}.  We approximate
the spectral energy distribution by a black body at a given
effective temperature $T_\eff$ and integrate over the {\it Kepler}
bandpass\hfil\break
(http://keplergo.arc.nasa.gov/kepler\_response\_hires1.txt)
and the 2MASS $H$-band bandpass respectively.  The latter is, of course,
not identical to the {\it WFIRST} bandpass, but for the reddish
stars we are considering, which peak in the {\it WFIRST} band and
well redward of the central response of the {\it Kepler} band, the 
main sensitivity to temperature is through the {\it Kepler} band.
Figure~\ref{fig:ampratio} shows the resulting ratio of amplitudes
$A_{H/K_p}$.  The main point is that it is essentially flat over
a broad range of temperatures.  Fitting the points to a parabola,
we obtain
\begin{equation}
A_{H/K_p} = 0.448 + 0.068(T_{5000}-1) + 0.074(T_{5000}-1)^2
\label{eqn:parab}
\end{equation}
where $T_{5000}\equiv T_\eff/5000\,$K. We use Equation~(\ref{eqn:parab})
in our simulations, although $A_H/Kp=0.45$ would be a very good approximation.

\begin{figure}[t!]
\begin{center}
\resizebox{85mm}{!}{\includegraphics{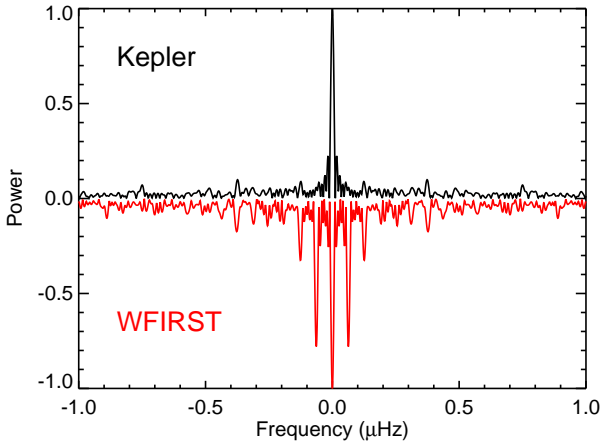}}
\caption{Spectral window function for a typical {\it Kepler} time series 
(top panel, black) and after degrading the time series to a typical duty cycle 
expected for {\it WFIRST} (bottom panel, red).  Note that while $\sim 80\%$
of the power is displaced to side lobes, these are still contained within
a $\sim 0.25\,\mu$Hz envelope, which is substantially narrower than
most spectral features of interest.}
\label{fig:window}
\end{center}
\end{figure}

\begin{figure}
\begin{center}
\resizebox{85mm}{!}{\includegraphics{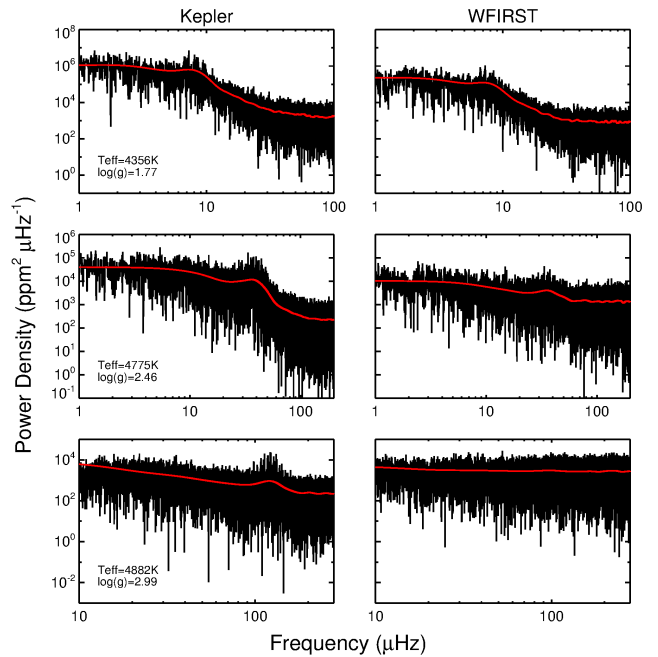}}
\caption{Power spectra of {\it Kepler} observations (left panels) and simulated 
{\it WFIRST} observations (right panels) for three red giants in 
different evolutionary stages: high-luminosity red giant 
(top panels), red clump (middle panels) and 
low-luminosity red giant (bottom panels). 
Red lines show the power spectra smoothed 
with a Gaussian with a full-width half-maximum of $2\Delta\nu$. Estimated 
stellar properties are given in the left panels, with a more complete
description given in Table~\ref{tab:params}.}
\label{fig:simplot}
\end{center}
\end{figure}

{\subsection{Spectral Frequency Response Function}
\label{sec:ofrf}}

The {\it WFIRST} microlensing observations will take place
in 72-day intervals centered on quadrature.  The exact scheduling
of the six campaigns has not been decided upon.  For simplicity,
we approximate that these take place over 3 consecutive years.
This schedule then mimics the typical day/night gaps of
ground-based observations (except that the cycle is 182 times
longer) and so can be expected to induce similar aliasing
(but with frequency spacings that are 182 times smaller).
This is illustrated by Figure~\ref{fig:window}, where we show
a typical {\it Kepler} spectral window function and the
result of imposing the {\it WFIRST} observing windows on 
the {\it Kepler} time series.  Note that while $\sim 80\%$ of
the power is now in sidelobes with spacing $\sim 0.06\,\mu$Hz
(corresponding to the 182 day on/off cycle), the entire
envelope of this ``distributed'' power is contained within
$\sim 0.25\,\mu$Hz.  We will see below that this has no
practical impact (except possibly for the very brightest stars)
because the frequency spacings that are
potentially measurable are much larger than this.  Hence,
when comparing {\it WFIRST} to {\it Kepler}, the main difference
is photometric noise, to which we now turn.

{\subsection{Photometric Noise Estimates}
\label{sec:photometric}}

As discussed in Section~\ref{sec:airy_phot}, the photometric noise
is function of apparent brightness, which is in turn determined by
three parameters: luminosity, distance, and extinction.  Here we focus
on stars in the Galactic bulge (distance modulus $\mu=14.7$), and
adopt a typical extinction for these fields, $A_H=0.5$.  Therefore,
$H = M_H + 15.2$.  We then find from Figure~\ref{fig:prec} that
\begin{equation}
\sigma_W = 1.0 \times 10^{(2/15)M_H}\,{\rm mmag}.
\label{eqn:sigmah}
\end{equation}

For each star considered, we evaluate 
$M_H = -2.5\log(L/L_\odot) + M_{\rm bol,\odot} - {\rm BC}_H$
where $M_{\rm bol,\odot} = 4.75$ and ${\rm BC}_H$ is the $H$-band
bolometric correction adopted from \citet{casagrande14} by interpolating
their tables in $T_\eff$, $\log g$, and [Fe/H].  Then taking account
of the fact that {\it WFIRST} observations occur twice as frequently as
{\it Kepler} observations, we create a simulated lightcurve
of flux in {\it WFIRST} $H$-band, $F_{H,i}$, by
\begin{equation}
F_{H,i} = (F_{Kp,i} - \overline{F_{Kp}})A_{H/Kp} + 
N\biggl(0,{\sigma_W\over \sqrt{2}}\biggr),
\label{eqn:simulh}
\end{equation}
where $F_{Kp,i}$ is the $i$th observed {\it Kepler} flux measurement,
$\overline{F_{Kp}}$ is the mean of these data points, and $N(q,\sigma)$
is a Gaussian random variable of mean $q$ and variance $\sigma^2$.

\begin{figure}
\begin{center}
\resizebox{85mm}{!}{\includegraphics{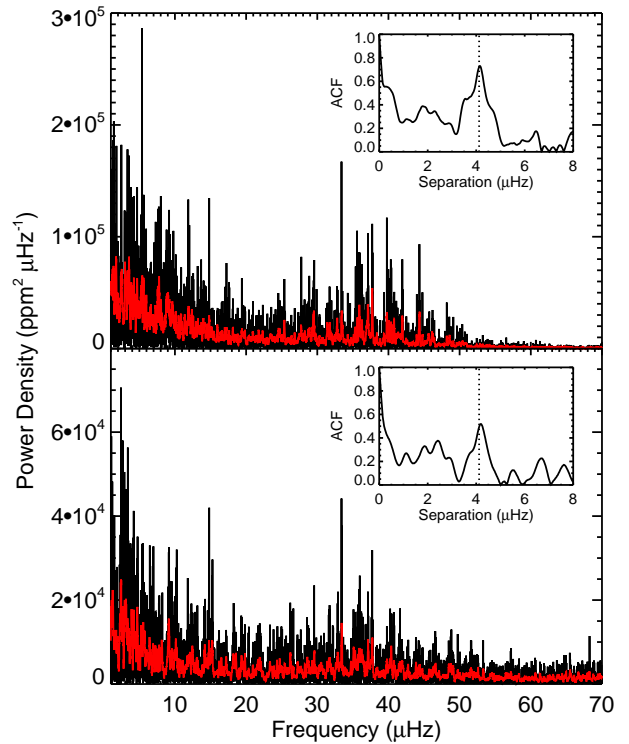}}
\caption{Oscillation spectrum for KIC\,2836038 as observed by {\it Kepler}
(top panel) and as simulated for {\it WFIRST} (bottom panel). 
Red lines are the power 
spectra smoothed with a boxcar width of $1\mu$Hz. The insets show the 
autocorrelation of the smoothed power 
spectrum between $20-60\mu$Hz after correcting 
the granulation background. The dotted line marks the published value for the 
large frequency separation $\Delta\nu$ \citep{apokasc}.}
\label{fig:simplot2}
\end{center}
\end{figure}

{\subsection{Simulated {\it WFIRST} Asteroseimology}
\label{sec:simast}}

To assess the potential of {\it WFIRST} asteroseimology, 
we simulate {\it WFIRST} observations of three red giants in different
evolutionary stages that were observed by {\it Kepler},
first ``placing them'' in the Galactic bulge, behind $A_H=0.5$ mag of
extinction (as described in Section~\ref{sec:photometric}).
These range from the high-luminosity red giant KIC 2437965 
($\nu_\max\sim 7.5\,\mu$Hz), through the clump giant KIC 2836038
($\nu_\max\sim 35.7\,\mu$Hz), to the low luminosity red giant KIC 6605620
($\nu_\max\sim 120.5\,\mu$Hz). See Table~\ref{tab:params}.

\begin{table}[t!]
\begin{center}
\begin{tabular}{l c c c}        
\hline         
Parameter & KIC\,2437965 & KIC\,2836038 & KIC\,6604620\\	
\hline		
$T_\eff$\ (K)     & 4356   & 4775   & 4882    \\
$\log g$\ (cgs)  & 1.765  & 2.460  & 2.993    \\
$\rm [Fe/H]$\ (dex)& 0.43 & 0.33   &-0.33    \\
$R (R_\odot)$     & 24.94  & 11.28  & 6.50    \\
$M_{H}$ (mag)    & -3.14   & -1.60  & -0.45    \\
$A_{H}/A_{Kp}$    & 0.44    & 0.45   & 0.45    \\
$\sigma_{W}$ (mmag)& 0.381 & 0.613  & 0.871    \\
Reference         &P14    & C14    & C14      \\
\hline		
\end{tabular} 
\caption{Fundamental properties and simulation parameters for
{\it WFIRST} simulations.  P14 = \citet{apokasc}, C14 = \citet{casa14}.} 
\label{tab:params}
\end{center} 
\end{table}

Figure \ref{fig:simplot} shows the power spectra for the original
{\it Kepler} data (left panels) and the simulated {\it WFIRST} data (right
panels). 
As expected from the well-known relation between oscillation
amplitudes and luminosity \citep{KB95,huber11b}, oscillations are more
easily detected in high-luminosity giants.
Despite the decreased oscillation amplitude and increased photon noise, 
the oscillations are clearly detectable for the high-luminosity red giant
and for the clump giant. For the low-luminosity red giant, both the sloping
background granulation and oscillation signals are completely masked by noise.
Note that even for the luminous giant, the hump of the oscillation
power excess (centered on $\nu_\max$) has a width that
is an order of magnitude larger than the envelope of the spectral
window shown in Figure~\ref{fig:window}.  This means that the {\it WFIRST}
semi-annual window function has no impact on detectability of this feature.

Figure \ref{fig:simplot2} shows a close-up of the power excess for the
red-clump star for {\it Kepler} data (top panel) and {\it WFIRST}
simulated data (bottom panel).  The {\it WFIRST} power spectrum is
clearly affected by the increased noise, which tends to bury
individual modes in the noise floor.  Note in particular that the
Gaussian noise that has been injected into the individual data points
results in a ``white noise'' floor in the Fourier transform, which is
absent from the {\it Kepler} spectrum for $\nu\ga 55\,\mu$Hz.  This
will make the extraction of individual frequencies challenging.
Importantly, however, the data have sufficiently high S/N to allow a
clear detection of $\langle\Delta\nu_{nl}\rangle$. This is
demonstrated in the insets, which show an autocorrelation of the
spectrum centered on the power excess. In both cases a clear peak is
visible near $4.1\,\mu$Hz, which agrees with the published
$\langle\Delta\nu_{nl}\rangle$ value \citep{apokasc}.  Because this
value is much larger than the $0.06\,\mu$Hz aliases in the {\it WFIRST}
spectral window function (induced by semi-annual gaps) seen in
Figure~\ref{fig:simplot}, these aliases do not critically impact
measurement of this feature.  However, for extremely bright stars
with $\sim 10$ times larger radii ($\sim 1000$ times lower gravity,
and so $\sim 30$ times smaller $\langle \nu_{nl}\rangle$), this would
become an issue (assuming that their $\sim 1\,$yr periods were adequately
resolved in the {\it WFIRST} mission).  Note that the
autocorrelations were calculated using power spectra that were
corrected for the granulation background as described by
\citet{huber09}.  While the exact detection limit will depend on the
actual photometric performance, these tests confirm that
{\it WFIRST} has great potential for asteroseismology of red giants above
and slightly below the red clump.

In evaluating this potential, it is important to keep in mind that
{\it WFIRST} astrometry will help distinguish genuine seismic
detections from instrumental power excess (e.g., Figure 6 
from \citealt{hekker11}), as well as noise spikes and
aliases. For example, luminosities based on {\it WFIRST} parallaxes will
provide a good initial guess for the location of the power excess due
to oscillations \citep{stello08}. Furthermore, {\it WFIRST} 
parallaxes (Section~\ref{sec:airy}) combined
with flux and surface-brightness measurements (Section~\ref{sec:angular})
will yield the radius $R$, and as outlined in Section~\ref{sec:intro},
this can be combined with the measurement of $\nu_\max$ (from the middle panel
of Figure~\ref{fig:simplot2}) to estimate the mass, even without a direct
measurement of $\langle\nu_{nl}\rangle$ (Equation~\ref{eqn:mass2}).
But this also means that the approximate location of 
$\langle\nu_{nl}\rangle$ can be determined from the measurement $\nu_\max$
and $R$, via
\begin{equation}
{\langle\Delta\nu_{nl}\rangle\over\langle\Delta\nu_{nl}\rangle_\odot}\simeq
\biggl({\nu_\max\over\nu_{\max,\odot}}\biggr)^{1/2}
\biggl({T_\eff\over T_{\eff,\odot}}\biggr)^{1/4}
\biggl({R\over R_\odot}\biggr)^{-1/2}.
\label{eqn:nupred}
\end{equation}
Then, once the location of this peak is approximately identified from
the measurements of $\nu_\max$ and $R$ (and these scaling relations),
it can be measured more precisely from the autocorrelation function.
A simpler version of this procedure is already in routine use (e.g.,
\citealt{huber09}) based on the purely asteroseismic 
empirical scaling relation of 
\citet{stello09},
\begin{equation}
{\langle\Delta\nu_{nl}\rangle\over \mu{\rm Hz}}
=(0.263\pm 0.009)\biggl({\nu_\max\over \mu{\rm Hz}}\biggr)^{0.772\pm 0.005}.
\label{eqn:stello}
\end{equation}
  However,
the additional radius information entering Equation~(\ref{eqn:nupred})
makes it more robust against unusual stars that systematically differ
from the local calibration sample underlying Equation~(\ref{eqn:stello}).
Such stars may be more common in the Galactic bulge than they are locally.
If the autocorrelation function is too noisy to identify a clear
peak even with this assist, 
or if the noise renders the error in the position of the peak too large to
be usable.
it will still be possible to estimate
the mass, just from the measurements of $\nu_\max$ and $R$ via 
Equation~(\ref{eqn:mass2}).

\begin{figure}
\includegraphics[width=85mm]{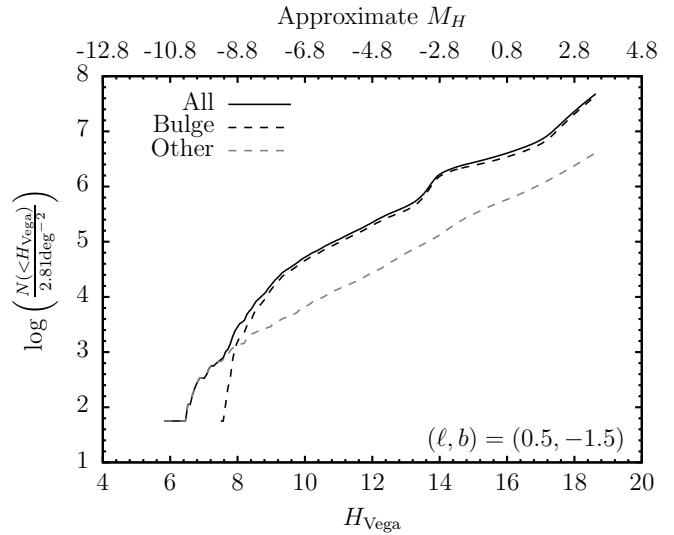}
\caption{Cumulative distribution of stars in a typical choice for the
$2.8\,{\rm deg}^2$ {\it WFIRST} field toward the Galactic bulge.
Solid curve represents all stars while dashed curve represents stars
lying in the Galactic bulge.  To understand anticipated
astroseismic performance of {\it WFIRST}, one should subtract
$\mu + A_H=15.2$ from the observed $H$ magnitude and compare with
the absolute magnitudes $M_H$ shown in Table~\ref{tab:params} as
displayed in Figure~\ref{fig:simplot}.  This is strictly true for
bulge stars and approximately true for the others, which are mainly
in the foreground disk.  {\it WFIRST} will return very good astroseismic
data for about 1 million stars.  Upper scale shows approximate absolute
magnitude, assuming Galactocentric distance and fiducial extinction.
}
\label{fig:cum}
\end{figure}

{\subsection{Availability of Asteroseismic Targets}
\label{sec:availability}}

Figure~\ref{fig:cum} shows the cumulative distribution of stars 
by $H$-band magnitude for a typical choice of the $2.8\,\rm deg^2$
{\it WFIRST} field, estimated using the TRILEGAL Galactic model 
\citep{vanhollebeke09,girardi12}.  The subset of these that lie in the Galactic
bulge are delineated by the bold dashed line.  The bump near $H\sim 13.5$
is the red clump.  The quality of the
asteroseismic data for the bulge stars can be directly evaluated
by comparison to the three stars shown in Figure~\ref{fig:simplot},
whose parameters are given in Table~\ref{tab:params}.  These
have $M_H=(-3.1,-1.6,-0.5)$, 
so in our approximate of uniform extinction, $H=(12.1,13.6,14.7)$.
The remaining stars are overwhelmingly in the foreground Galactic disk
and so are dimmer at fixed apparent magnitude.  Thus, at the same
$H$ magnitude they will, of course, have the same photometric errors,
but being dimmer, their asteroseismic signals will generally be
somewhat weaker.  Nevertheless, because most of these stars
lie within a magnitude of the distance modulus of the bulge, 
Figure~\ref{fig:simplot} provides a good qualitative indicator
of astroseismic performance for these stars as well as a function 
of magnitude.  

Thus, in total, there will be roughly 1 million stars with $H<14$ with
good astroseismic solutions, particular when account is taken of
the precision parallaxes (and so radii) for these stars from {\it WFIRST}
astrometry.  In particular, for $H=14$, these will have 
$\sigma(\pi)=0.34\,\muas$ parallaxes, corresponding to 0.3\% distance
errors.

\acknowledgments

Work by AG was supported by NSF grant AST 1103471 and
NASA grant NNX12AB99G.
MP acknowledges support by The Thomas Jefferson
Chair for Discovery and Space Exploration.
DH acknowledges support by NASA Grant NNX14AB92G issued through the
Kepler Participating Scientist Program.


\end{document}